\documentstyle[aps,prb,multicol,epsfig]{revtex}

\newcommand{\ic}        { {\rm i}}
\newcommand{\iz}        { \ic 0^{+}}
\newcommand{\order}[1]  { {{\cal O}\left(#1\right)}}

\newcommand{\abs}[1]    { {\left| #1 \right|}}
\newcommand{\real}      { {\rm Re\,} }
\newcommand{\imag}      { {\rm Im\,} }
\newcommand{\sign}[1]   { {\rm sgn}(#1) }

\newcommand{\vect}[1]   { {\mathbf #1 } }
\newcommand{\vk}        { {\vect{k}} }

\newcommand{\mean}[1]   { \left<  \, {#1} \, \right> }

\newcommand{\Gi}[1]     { G_{i} }
\newcommand{\Di}[1]     { D_{i} }
\newcommand{\PiZch}[1]	{ ^{0}\Pi }
\newcommand{\Si}[1]	{ \Sigma_{i} }

\newcommand{\Gisig}     { \Gi{\sigma} }
\newcommand{\GiFree}[1] { \Gi{#1}^{0} }
\newcommand{\GiFsig}    { \GiFree{\sigma} }

\newcommand{\Disig}     { \Di{\sigma} }
\newcommand{\DiFree}[1] { \Di{#1}^{0} }
\newcommand{\DiFsig}    { \DiFree{\sigma} }

\newcommand{\ReSi}[1]   { \Si{#1}^{\rm R} }
\newcommand{\ImSi}[1]   { \Si{#1}^{\rm I} }

\newcommand{\SiPT}[1]   { \Si{#1} }
\newcommand{\ReSiPT}[1] { \SiPT{#1}^{\rm R} }
\newcommand{\ImSiPT}[1] { \SiPT{#1}^{\rm I} }

\newcommand{\omo}       { \omega_{0}}
\newcommand{\omtilde}   { \tilde{\omega} }

\newcommand{\ReDelta}   { \Delta_{{\rm R}} }
\newcommand{\ImDelta}   { \Delta_{{\rm I}} }
\newcommand{\DelZ}      { \Delta_{0} }

\newcommand{\DelZD}     { \Delta_{0} D }
\newcommand{\DelZZ}     { \Delta_{0} Z }
\newcommand{\sDelZD}    { \sqrt{\Delta_{0} D} }

\newcommand{\bibart}[5]         {{#1}, {#2} {\bf #3}, {#4} {(#5)}}

\newcommand{\bibbook}[4]        {{#1}, {\em #2}, #3 (#4)}

\newlength{\Units}

\begin{document}
\draft

\title{Spectral properties of a narrow-band Anderson model} 

\author{Steffen Sch\"afer and David E Logan}
\address{Oxford University, Physical and Theoretical Chemistry
Laboratory, South Parks Road, Oxford OX1 3QZ, UK}

\date{August 16, 2000; revised October 16, 2000}

\maketitle

\begin{abstract}

We consider single-particle spectra of a symmetric narrow-band
Anderson impurity model, where the host bandwidth $D$ is small
compared to the hybridization strength $\DelZ$. Simple 2nd order
perturbation theory (2PT) in $U$ is found to produce a rich spectral
structure, that leads to rather good agreement with extant Lanczos
results and offers a transparent picture of the underlying physics. It
also leads naturally to two distinct regimes of spectral behaviour,
$\DelZZ/D\gg 1 $ and $\ll 1$ (with $Z$ the quasi-particle weight),
whose existence and essential characteristics are discussed and shown
to be independent of 2PT itself. The self-energy $\Si{\sigma}(\omega)$
is also examined beyond the confines of PT. It is argued that on
frequency scales of order $\omega\sim\sDelZD$, the self-energy in {\em
strong} coupling is given precisely by the 2PT result, and we point
out that the resultant poles in $\SiPT{\sigma}(\omega)$ connect
continuously to that characteristic of the atomic limit. This in turn
offers a natural rationale for the known inability of the skeleton
expansion to capture such behaviour, and points to the intrinsic
dangers of partial infinite-order summations that are based on PT in
$U$.

\end{abstract}

% %%%%%%%%%%%%%%%%%%%%%%%%%%%%%%%%%%%%%%%%%%%%%%%%%%
% PACS numbers according to 1999 PACS
% %%%%%%%%%%%%%%%%%%%%%%%%%%%%%%%%%%%%%%%%%%%%%%%%%%

% other possibilities are:
% 71.10.Fd Lattice fermion models (Hubbard model, etc.);
% 71.55.-i Impurity and defect levels;

\begin{multicols}{2}
\narrowtext

% %%%%%%%%%%%%%%%%%%%%%%%%%%%%%%%%%%%%%%%%%%%%%%%%%%%%%%%%%%%%%%%%%%%%%%

\section{Introduction}

Since its inception nearly forty years ago, the Anderson impurity
model (AIM) \cite{Anderson61} has become one of the generic models of
correlated electron physics, permitting at possibly the simplest level
a detailed study of the competition between local interactions and
electron itinerancy. The great majority of previous work, reviewed
comprehensively in Ref.~\onlinecite{Hewson}, has focused on an
impurity coupled to a conduction band whose width ($D$) represents a
large energy scale --- large compared to the one-electron
hybridization strength ($\DelZ$) that characterizes the host/impurity
coupling. The opposite extreme of a {\em narrow-band} AIM has however
been invoked \cite{Hofstetter99,Kehrein98} in relation to the Hubbard
model in large dimensions \cite{Vollhardt}, where the correlated
lattice problem maps onto an effective impurity model coupled to a
self-consistently determined bath \cite{Georges96}. Here, in the
vicinity of the Mott transition occurring in the particle-hole
symmetric ($\frac{1}{2}$-filled) Hubbard model, an effective
narrow-band AIM is generated self-consistently in the now widely
accepted scenario for the transition \cite{Georges96} that arose
originally from the iterated perturbation theory (IPT) approach of
Kotliar, Georges and coworkers \cite{Georges92,IPT}.

Recently Hofstetter and Kehrein \cite{Hofstetter99} have undertaken a
numerical study of the symmetric narrow-band AIM, freed from the
additional complications of self-consistency and itself defined by
$\DelZ\gg D$.  Thermodynamic and spectral properties of the model were
found thereby to exhibit a range of behaviour that is quite distinct
from the conventional \cite{Hofstetter99} wide-band
AIM. Single-particle spectra for example, determined via finite-size
Lanczos calculations, show a much richer structure than their
wide-band counterpart. At low frequencies in particular they exhibit
an intricate competition between various physical effects and their
associated energy scales, reflecting ultimately the importance in the
narrow-band AIM of an energy scale $\omo\sim\sDelZD$ that while large
compared to the host bandwidth $D$ is typically small compared to the
on-site interaction $U$.

Spectral properties of the narrow-band AIM are discussed in the
present paper, a complementary analytical study based in part upon 2nd
order perturbation theory (2PT) in $U$. This is found to provide a
qualitatively correct description on all energy scales, suggests a
simple physical interpretation of the most prominent spectral
features, and produces rather good agreement with the Lanczos results
of Hofstetter and Kehrein \cite{Hofstetter99}. We also show,
independent of the intrinsic limitations of 2PT, that spectral
behaviour divides into two distinct regimes that are borne out by the
calculations of Ref.~\onlinecite{Hofstetter99}: (a) weak-coupling for
$\DelZZ/D \gg 1$ (with $Z$ the quasi-particle weight), and (b) a
strong coupling (Kondo) regime for $\DelZZ/D \ll 1$.

For the weak to moderate interaction strengths reliably accessible by
Lanczos calculations, Hofstetter and Kehrein
\cite{Hofstetter99,Kehrein98} have shown that pole contributions to
the interaction self-energy $\Si{\sigma}(\omega)$, occurring on the
scale $\omo\sim\sDelZD$, are an integral facet of the narrow-band AIM;
but that they cannot be captured to any order in a skeleton expansion.
Within 2PT, $\SiPT{\sigma}(\omega)$ is indeed found to be dominated
entirely by poles on this order. However in strong coupling
$U\gg\DelZ$, 2PT will certainly be qualitatively deficient on the
lowest energy scales $\abs{\omega}\ll D$ characteristic of the Kondo
resonance.  We show nonetheless that in strong coupling the pole
contributions to the self-energy are again given {\em precisely} by
the result arising from 2PT. These poles are thus a ubiquitously
dominant and essentially $U$-independent characteristic of the
narrow-band AIM. They are moreover continuously connected to the
atomic limit of the model, which sheds light on the inability of a
skeleton expansion to recover them, and points to the intrinsic
dangers of partial infinite-order summations based on PT in $U$.

The Hamiltonian for the AIM is given in standard notation by
\begin{eqnarray}
\label{Hamiltonian}
\hat{H} & = & 
\sum\limits_{\vk\sigma}\varepsilon_{\vk}\hat{n}_{\vk\sigma}
\,+\,
\sum\limits_{\sigma} \varepsilon_{i}\hat{n}_{i\sigma} 
\,+\,
U\hat{n}_{i\uparrow} \hat{n}_{i\downarrow} 
\nonumber \\ &&
\,+\,
\sum\limits_{\vk\sigma} 
\left( V_{i\vk} c^{+}_{i\sigma} c_{\vk\sigma} + {\mathrm h.c.} \right)
\end{eqnarray}
where the first term describes electrons (of spin
$\sigma=\uparrow,\downarrow$) in a metallic host band of dispersion
$\varepsilon_{\vk}$. The following two terms refer to the impurity,
with $\varepsilon_{i}$ the impurity level and $U$ the on-site Coulomb
interaction. The final term describes the one-electron hybridization
between the impurity and host. Throughout this article, as in
Ref.~\onlinecite{Hofstetter99}, we study the particle-hole symmetric
AIM where $\varepsilon_{i}=-U/2$. In this case, for all interaction
strengths, the Fermi level remains fixed at its non-interacting value
and the impurity charge
$n_{i}=\mean{\hat{n}_{i\uparrow}+\hat{n}_{i\downarrow}}=1$;
single-particle spectra are thus symmetric with respect to the Fermi
level, $\omega =0$.

\section{Non-interacting limit}
\label{sec:U0}

We first consider briefly the noninteracting problem
$\varepsilon_{i}=-U/2=0$ which, although rather trivial by itself,
generates two inherent energy scales that control the subsequent
evolution of single-particle properties with interaction strength. For
$U=0$ the (causal) single-particle impurity Green function
$
\GiFsig(\omega)=\real\GiFsig(\omega)-\ic\,\sign{\omega}\pi\DiFsig(\omega)
$
is given by \cite{Hewson}
\begin{equation}
\label{GFree}
\GiFsig(\omega) \,=\, \left[z-\Delta(\omega)\right]^{-1}
\end{equation}
with $z=\omega+\iz\sign{\omega}$ and
$\Delta(\omega)=\ReDelta(\omega)-\ic\,\sign{\omega}\ImDelta(\omega)$
the hybridization function, 
$
\Delta(\omega)= 
\sum_{\vk} {\abs{V_{i\vk}}^{2}}/{\left(z-\varepsilon_{\vk}\right)}
$.

We are interested in the {\em narrow-band case}, meaning
$D\gg\DelZ=\ImDelta(\omega=0)$ with $D$ the host bandwidth.  The
physics of the model is naturally rather insensitive to the precise
form of the hybridization. We thus take $\ImDelta(\omega)$ to consist
of a single flat band of intensity $\DelZ$, ranging from $-D$ to $+D$;
so that in total
\begin{equation}
\label{nbhybrid}
\Delta(\omega) \,=\, 
\frac{\DelZ}{\pi} \, \log{\abs{\frac{\omega+D}{\omega-D}}} 
\,-\,
\ic \DelZ\,\sign{\omega}\,\theta(D-\abs{\omega})
\end{equation}
where $\ReDelta(\omega)$ follows via Hilbert transformation.

The noninteracting single-particle spectrum consists of two
contributions: (i) a low-energy continuum for $\omega\in [-D,+D]$,
arising from the non-zero imaginary part of the hybridization; (ii)
two poles, one lying above and the other one symmetrically below the
continuum. The fraction of spectral weight residing in the poles
depends strongly on the bandwidth of the host metal.  In the usual
wide-band model, $D\gg\DelZ$, most of the spectral weight is
concentrated in the single-particle band, while the poles are
exponentially weak and hence irrelevant in practice.

The situation is however radically different in the present
narrow-band model, defined by $D\ll\DelZ$. Here the poles are the most
prominent features of the spectrum, suggesting that the narrow host
behaves in effect as a single site or level
\cite{Hofstetter99}. As this ``host site'' is coupled to the impurity,
a bonding and an anti-bonding orbital form in analogy to the ${\rm
H}_{2}$ molecule. These orbitals occur at frequencies $\pm\omo$ far
outside the band, obtained by expanding
$\ReDelta(\omega)\sim(2/\pi)\DelZD/\omega$ in the denominator of
(\ref{GFree}) to yield
\begin{equation}
\label{omega0}
\omo\,=\,\sqrt{\frac{2\DelZD}{\pi}}
\;\mbox{.}
\end{equation}
The pole-weight of each orbital is $q=(2\DelZ-\pi D)/(4\DelZ-\pi
D)\sim\frac{1}{2}+\order{D/\DelZ}$, showing that they carry almost the
entire spectral density.

The noninteracting single-particle spectrum is thus characterized by a
weak continuum of weight $\order{D/\DelZ}$, ranging over the lowest
energy scale $\abs{\omega}\le D$, and two molecular orbitals carrying
most of the spectral intensity.  The latter occur at the orbital
bonding energy $\omo$ which, although much larger than $D$, defines a
second low-energy scale $\omo\sim\sDelZD\ll\DelZ$.  Central questions
then arising include: how does this low-energy scale evolve with
increasing interaction strength, over what range of $U$ does it
constitute the dominant low-energy physics of the problem, and when by
contrast is the latter dominated by the Kondo effect that one expects
to prevail for sufficiently large interactions? It is these questions
we aim to address in the following sections.

\section{Perturbation theory}
\label{sec:2PT}

The impurity Green function for the interacting system,
$\Gisig(\omega)$, is related to $\GiFsig(\omega)$ by the Dyson
equation
\begin{equation}
\label{Dyson}
\Gisig(\omega) \,=\, 
\frac{1}{\left[\GiFsig(\omega)\right]^{-1}-\Si{\sigma}(\omega)}
\;\mbox{.}
\end{equation}
This simply defines the interaction self-energy $\Si{\sigma}(\omega)=
\ReSi{\sigma}(\omega)-\ic\,\sign{\omega}\ImSi{\sigma}(\omega)$,
considered here via perturbation theory (PT) in $U$ about the
noninteracting limit. One does not doubt such an approach in principle
for, although we are interested in the narrow-band regime, the
impurity is coupled to a {\em metallic} host and one thus anticipates
ubiquitously a Fermi liquid ground state that is perturbatively
connected to the noninteracting limit \cite{Hewson}. PT is of course
limited in practice, by the order to which it is taken.

The 1st order (Hartree) contribution to $\Si{\sigma}$, of $\frac{U}{2}
n_{i}=U/2$, is cancelled precisely by the bare site-energy
$\varepsilon_{i}=-U/2$. The lowest order process is thus of 2nd order
in $U$, whereby a propagating $\sigma$-spin electron generates a
particle-hole fluctuation of opposite spin on the impurity.
Diagrammatically, it reads
\begin{equation}
  \label{fig:Sigma2PT}
  \leavevmode
  \unitlength=2ex 
  \setlength{\Units}{\the\unitlength} 
  \begin{picture}(6,4)
  \put(0,2.0){\makebox(0,0)[c]{$\SiPT{\sigma}(\omega) \, \sim \,$}}
  \put(3,0.0){\epsfig{file=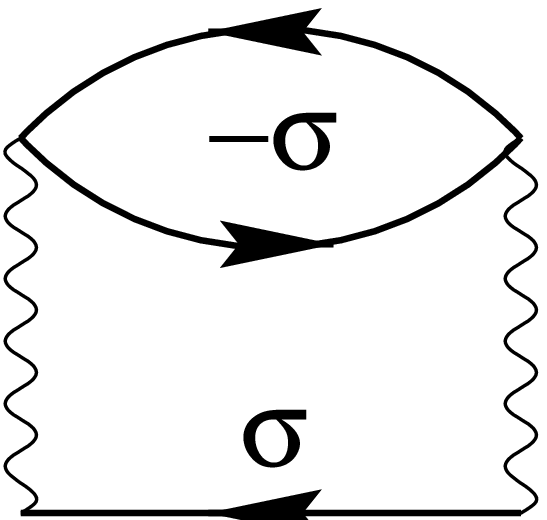,width=5\Units}}
  \end{picture}
\end{equation}
with wavy lines representing the interaction $U$ and left-going
(right-going) solid lines representing a free particle (hole)
propagator $\GiFsig$.

\subsection{Particle-hole polarisation propagator}

The $-\sigma$-spin particle-hole bubble appearing in
(\ref{fig:Sigma2PT}) stands for the noninteracting polarisation
propagator
\begin{equation}
\label{Pi0}
\PiZch{-\sigma}(\omega) \,=\, \ic 
\int\limits_{-\infty}^{+\infty} \frac{{\rm d}\omega'}{2\pi}\, 
\GiFree{-\sigma}(\omega')\,\GiFree{-\sigma}(\omega'-\omega)
\; \mbox{.}
\end{equation}
Using the Hilbert transform relation for $\GiFree{-\sigma}(\omega)$, a
short calculation gives
\begin{eqnarray}
\label{ImPi0}
\frac{1}{\pi} \imag\PiZch{-\sigma}(\omega)   & = &  
\theta(\omega)\int\limits_{0}^{+\abs{\omega}} {\rm d}\omega'\,
\DiFree{-\sigma}(\omega')\,\DiFree{-\sigma}(\omega'-\omega) 
\nonumber \\
& & + 
\theta(-\omega)\int\limits_{-\abs{\omega}}^{0} {\rm d}\omega'\,
\DiFree{-\sigma}(\omega')\,\DiFree{-\sigma}(\omega'-\omega) 
\; \mbox{.}
\end{eqnarray}
Since $\imag\PiZch{-\sigma}(\omega)$ is symmetric with respect to the
Fermi level $\omega=0$, it is sufficient to discuss the first term
($\omega>0$). It describes particle-hole excitations obtained by
transferring an electron from an occupied state below the Fermi level
($\omega'-\omega<0$) to an unoccupied state above it ($\omega'>0$).
The narrow-band model is dominated by particle-hole excitations
between the bonding and the anti-bonding orbital, involving an energy
cost of $2\omo$, which in $\frac{1}{\pi}\imag\PiZch{-\sigma}(\omega)$
translates to pole of weight $\simeq 1/4$ at $\omega=2\omo$.

In addition, $\imag\PiZch{-\sigma}(\omega)$ contains two continua,
associated with particle-hole excitations to/from the single-particle
band and therefore of small spectral weight. A first continuum arises
for $\omega\in [\omo,\omo+D]$ (net weight $\order{D/\DelZ}$),
corresponding to electronic transfer between the single-particle band
and one of the orbitals. A second appears for low frequencies
$\abs{\omega}\le 2D$ (net weight $\order{[D/\DelZ]^{2}}$), arising
from particle-hole excitations within the single-particle band. As
$\DiFree{-\sigma}(\omega)$ is finite at the Fermi level $\omega=0$,
the low-frequency behaviour of the latter contribution is
\begin{equation}
\label{ImPi0_omega0}
\frac{1}{\pi}\imag\PiZch{-\sigma}(\omega) \, \stackrel{\omega\to 0}{\sim} \,
\abs{\omega} \left[ \DiFree{-\sigma}(0) \right]^{2}
\end{equation}
(which is used below to prove Fermi-liquid properties on the lowest
energy-scale).

\subsection{Self-energy in 2PT}
\label{sec:Sigma2PT}

The self-energy in 2nd order perturbation theory (2PT) is given by
(\ref{fig:Sigma2PT}), which translates to:
\begin{equation}
\label{Sigma2PT}
\SiPT{\sigma}(\omega) \,=\, U^{2}
\int\limits_{-\infty}^{+\infty} \frac{{\rm d}\omega'}{2\pi\ic}\, 
\PiZch{-\sigma}(\omega')\,\GiFree{\sigma}(\omega-\omega')
\end{equation}
In analogy to the above calculation for the polarisation propagator,
we use the Hilbert transforms for $\PiZch{-\sigma}$ and
$\GiFree{\sigma}$ to obtain:
\begin{eqnarray}
\label{ImSigmaPT2}
\ImSiPT{\sigma}(\omega) & = & 
\theta(\omega)\,U^{2}\,\int\limits_{0}^{+\abs{\omega}} {\rm d}\omega'\,
\imag\PiZch{-\sigma}(\omega')\,\DiFree{\sigma}(\omega-\omega') 
\nonumber \\
& & 
+\theta(-\omega)\,U^{2}\,\int\limits_{-\abs{\omega}}^{0} {\rm d}\omega'\,
\imag\PiZch{-\sigma}(\omega')\,\DiFree{\sigma}(\omega-\omega') 
\end{eqnarray}
Again, due to particle-hole symmetry, we may restrict the following
discussion to positive frequencies; and $\ReSiPT{\sigma}(\omega)$
follows from $\ImSiPT{\sigma}(\omega)$ by Hilbert transformation.

In the narrow-band limit, the 2PT self-energy is dominated by the
following process: an extra $\sigma$-electron is placed in the empty
anti-bonding orbital (energy cost $\omo$) and excites an electron of
opposite spin from the bonding to the anti-bonding orbital (energy
cost $2\omo$).  In $\ImSiPT{\sigma}(\omega)$, this generates a pole of
weight $\simeq \pi U^{2}/8$ at $\omega=3\omo$, that gives by far the
dominant contribution. Much weaker contributions to
$\ImSiPT{\sigma}(\omega)$ arise in addition from the extra electron
being placed in the single-particle band instead of the orbital, or
involve one of the particle-hole continua discussed above.  This
yields self-energy continua for
$\omega\in [2\omo,2\omo+D]$ (net weight $\order{U^{2} D/\DelZ}$),
$\omega\in[\omo,\omo+2 D]$ (net weight $\order{U^{2} [D/\DelZ]^{2}}$), and
$\abs{\omega}\le 3D$ with net weight $\order{U^{2} [D/\DelZ]^{3}}$. 
The latter results from a convolution, as in (\ref{ImSigmaPT2}), of
the low-frequency continuum of
$\imag\PiZch{-\sigma}(\omega)\propto\omega$ (see
eq.~(\ref{ImPi0_omega0})) and the single-particle band. It leads to
$\ImSiPT{\sigma}(\omega)\propto\omega^{2}$ close to the Fermi level
$\omega=0$, corresponding to conventional Fermi liquid behaviour
\cite{Luttinger61}.

\subsection{Spectral evolution in the narrow-band regime.}
\label{sec:spectra}

Numerical results for the impurity spectrum
$\Disig(\omega)=-\frac{1}{\pi}\sign{\omega}\imag\Gisig(\omega)$ within
2PT will be given in the following section. Here we develop an
approximate, but rather accurate analysis that shows in simple
physical terms how the spectrum evolves with interaction strength,
particularly on the low-energy scales of primary interest.

As discussed above the self-energy is dominated by poles at
$\omega=\pm 3\omo$, its continua being negligibly small in the
narrow-band regime and vanishing as $D\to 0$. We thus retain solely
the pole contributions, {\sl i.e.}
\begin{eqnarray}
\label{ReSigmaPT2}
\ReSiPT{\sigma}(\omega) &\simeq &
\frac{U^2}{8}\,\left[\frac{1}{\omega-3\omo}+\frac{1}{\omega+3\omo}\right]
\nonumber \\
& = & 
%\,=\,
\frac{U^2}{4}\,\frac{\omega}{\omega^2-9\omo^2}
\end{eqnarray}
with a corresponding quasi-particle weight
$Z=[1-(\partial\ReSi{\sigma}(\omega)/\partial\omega)_{\omega=0}]^{-1}$
given by 
\begin{equation}
\label{Z2PT}
Z \,=\,
\left[ 
1 \,+\,\left(\frac{U}{6\omo}\right)^{2}
\right]^{-1}
\;\mbox{.}
\end{equation} 
The evolution of $Z=Z(U)$ is thus controlled simply by the ratio
$U/\omo$, and for $U\gg\omo\sim\sDelZD$ in particular,
\begin{equation}
\label{Z2PT_lrgU}
Z \; \sim \; \left(\frac{6\omo}{U}\right)^{2}
\; = \; \frac{72}{\pi}\frac{\DelZD}{U^{2}}
\;\mbox{.}
\end{equation}

With $\SiPT{\sigma}(\omega)$ given from (\ref{ReSigmaPT2}),
$\Gisig(\omega)$ follows directly from (\ref{Dyson}). The resultant
spectrum then consists of a continuum for $\abs{\omega}<D$, and two
sets of poles at $\abs{\omega}=\omega_{+}$ and $\omega_{-}$ (``set''
meaning a pole at positive frequency and its counterpart occurring
symmetrically below the Fermi level $\omega=0$). The latter follow
from the single-particle pole equation
$\omega-\ReDelta(\omega)-\ReSiPT{\sigma}(\omega)=0$; insertion of
(\ref{ReSigmaPT2}) into which, together with the expansion
$\ReDelta(\omega)\sim\omo^2/\omega$ --- which is valid for
$\abs{\omega}\gg D$ --- yields a biquadratic with solutions
\begin{equation}
\label{omegapm}
{\omega_{\pm}} \, = \,
\frac{1}{4}\,\left[
\sqrt{U^2+\left(8\omo\right)^{2}}
\, \pm \,
\sqrt{U^2+\left(4\omo\right)^{2}}
\right]
\; \mbox{.}
\end{equation}
The set of low-frequency poles at $\omega_{-}\le\omo$, with pole
weights $q_{-}$, have their origin in the molecular orbitals of the
non-interacting problem, to which they reduce for $U/\omo\to 0$ where
$\omega_{-}\to\omo$ and $q_{-}\to1/2$.
In addition, for any $U>0$, a second set of poles arise at
$\omega_{+}\ge 3\omo$. In weak coupling $U/\omo\ll 1$, the latter have
negligible pole-weight, $q_{+}\sim (U/\omo)^{2}$. With increasing $U$
however, they drain weight from the low-frequency orbital remnants and
shift to progressively higher frequencies until, for
$U\gg\omo\sim\sDelZD$, they correspond to the Hubbard satellites and
overwhelmingly dominate the spectrum: $\omega_{+}\sim U/2$ with
pole-weight $q_{+}\sim 1/2-\order{[\omo/U]^{2}}$.

The above pole analysis is exact in the strict limit of vanishing
bandwidth $D\to 0$, for fixed $\omo\sim\sDelZD$. Here, the host
behaves precisely as a single orbital at the Fermi level $\omega=0$,
and the problem reduces to the simple two-site model discussed by
Lange \cite{Lange98} and employed by Hofstetter and Kehrein
\cite{Hofstetter99} to rationalize their Lanczos-determined spectra
for the narrow-band AIM. In this limit the hybridization
$\ImDelta(\omega)$ is a $\delta$-function, and its weight
\begin{equation}
\label{delta_hybrid}
\int\limits_{-\infty}^{+\infty} {\rm d}\omega \, \ImDelta(\omega)
\,=\,
2\DelZD
\,=\,
\pi\abs{V_{i0}}^{2}
\end{equation}
relates the bonding energy $\omo$ (eq.~(\ref{omega0})) to the hopping
between host and impurity, $\omo=V_{i0}$. As pointed out by Lange
\cite{Lange98}, 2PT is in fact exact in this case, the 4-pole spectrum
being given precisely by eq.~(\ref{omegapm}); for $U\gg\omo$ in
particular, the low-energy pole $\omega_{-}$ reduces to 
\begin{equation}
\label{omegam_medU}
\omega_{-} 
\; \sim \; \frac{6\omo^2}{U} 
\; = \; \frac{12\DelZD}{\pi U} \;=\; \frac{3}{2}\, J
\end{equation}
where $J=4 V_{i0}^{2}/U$ is the antiferromagnetic exchange coupling
between the impurity and the host orbital.

As a ``zero-bandwidth'' approximation, the two-site limit just
discussed is however rather limited as a means to interpret the
narrow-band AIM. It is certainly accurate for the high-energy spectral
poles $\omega_{+}$ (eq.~(\ref{omegapm})) that evolve for $U\gg\omo$ to
the Hubbard satellites at $\omega_{+}\sim U/2$. But it naturally fails
both on the lowest energy scale of the bandwidth $D$ and, relatedly,
in capturing the strong coupling behaviour of the low-energy spectral
poles at $\abs{\omega}=\omega_{-}$: the latter, by definition, can
only occur outside the band ($\omega_{-}>D$), whence
eq.~(\ref{omegam_medU}) suggests a breakdown of the two-site analogy
for $J\sim\order{D}$ or $U\sim\order{\DelZ}$.

It is however straightforward to obtain, for all interaction
strengths, a uniform description of the spectrum at low energies;
where by low energy we mean $\omega\lesssim\omo\sim\sDelZD$, thus
encompassing both the lower-energy poles $\omega_{-}$ and the
continuum for $\abs{\omega}<D$. For provided only that
$\abs{\omega}\ll 3\omo$, the self-energy given from (\ref{ReSigmaPT2})
reduces to its leading low-frequency expansion
$\ReSiPT{\sigma}(\omega)\sim -(1/Z-1)\omega$. Hence, for
$\abs{\omega}\ll 3\omo$, the impurity Green function $\Gisig(\omega)$
is given from (\ref{Dyson}), using (\ref{nbhybrid}) for the
hybridization, by
\begin{eqnarray}
\label{Glowfreq}
\lefteqn{\DelZ\Gisig(\omega)\,=\,}
\\
&&
\left[ \frac{D}{\DelZZ}\omtilde
\,-\,
\frac{1}{\pi}\log{\abs{\frac{1+\omtilde}{1-\omtilde}}}
\,+\,
\ic\,\sign{\omtilde}\,\theta(1-\abs{\omtilde})
\right]^{-1}
\nonumber
\end{eqnarray}
where $\omtilde=\omega/D$. This equation may be used to understand the
low-energy spectrum for all $U$. It depends upon the single parameter
$\DelZZ/D$, where the $U$-dependence of the quasi-particle weight $Z$
is given explicitly by eq.~(\ref{Z2PT}) for 2PT. And the
characteristic spectra divide into two distinct regimes, according to
whether $\DelZZ/D\gg 1$ or $\ll 1$, as now considered.

\subsubsection{$\DelZZ/D\gg 1$.}

From eq.~(\ref{Z2PT}), $\DelZZ/D\gg 1$ corresponds to $U/\DelZ\ll
6\sqrt{2/\pi}\simeq 5$. Here the spectral poles, at
$\abs{\omega}=\omega_{-}$ occur far outside the band. Specifically,
expanding $\log{\abs{\frac{1+\omtilde}{1-\omtilde}}}\sim 2/\omtilde$
for $\abs{\omtilde}\gg 1$, eq.~(\ref{Glowfreq}) yields
$\omtilde_{-}=\sqrt{\frac{2}{\pi}\frac{\DelZZ}{D}}\gg 1$, {\sl i.e.}
\begin{mathletters}
\label{polem_medU_Z}
\begin{equation}
\label{omegam_medU_Z}
\omega_{-} \, = \, \omo\sqrt{Z} 
\end{equation}
with pole-weight 
\begin{equation}
\label{weight_medU_Z}
q_{-} \,=\, \frac{Z}{2}
\;\mbox{.}
\end{equation}
\end{mathletters}
This is essentially the two-site limit result \cite{Lange98} for
$\omega_{-}$ --- eq.~(\ref{omegam_medU_Z}) with $Z$ from (\ref{Z2PT})
differs negligibly from eq.~(\ref{omegapm}) --- encompassing both weak
coupling $U/\omo\to 0$ where $Z\to 1$ and $\omega_{-}\to\omo$; as well
as $U\gg\omo\sim\sDelZD$ (subject to $U/\DelZ\ll 5$), where
$\omega_{-}$ reduces to eq.~(\ref{omegam_medU}). In addition of
course, eq.~(\ref{Glowfreq}) yields correctly a spectral continuum for
$\abs{\omtilde}=\abs{\omega}/D<1$. But since $\frac{D}{\DelZ
Z}\abs{\omtilde}\ll 1$ for all $\abs{\omtilde}<1$, the continuum
reduces to that characteristic of the non-interacting limit, $U=0$
(where $\frac{D}{\DelZ}\omtilde$ in eq.~(\ref{Glowfreq}) is likewise
negligible for the narrow-band model).

Hence, for $\DelZZ\gg D$ ({\sl i.e.} $U/\DelZ\ll 5$), the spectral
'action' takes place exclusively in the poles $\omega_{-}$ (and
$\omega_{+}$) and is well captured by the two-site limit
\cite{Hofstetter99,Lange98}, while the low-energy continuum is
essentially unaffected by electron interactions.

\subsubsection{$\DelZZ/D\ll 1$.}

The above picture is marked contrast to that arising from $\DelZZ\ll
D$ ({\sl i.e.} $U/\DelZ\gg 5$), where the $\omega_{-}$ poles are
exponentially close to the band edge, being given from
(\ref{Glowfreq}) by
\begin{mathletters}
\begin{equation}
\label{omegam_lrgU}
\frac{\omega_{-}}{D} \;=\; 
1\,+\,2 \exp\left(-\frac{\pi D}{\DelZZ}\right)
\;\mbox{.}
\end{equation}
The corresponding pole-weight is however exponentially small,
\begin{equation}
\label{weight_lrgU}
q_{-} \;=\; \frac{2\pi D}{\DelZ}
\exp\left(-\frac{\pi D}{\DelZZ}\right)
\end{equation}
\end{mathletters}
so the low-energy poles in effect are irrelevant.

All the action now takes place in the continuum: from
(\ref{Glowfreq}), and for all $\abs{\omtilde}<1$ save exponentially
close to the band edges, the spectrum for $\DelZZ\ll D$ is given by
\begin{equation}
\label{Di_lrgU}
\pi\DelZ\Disig(\omega)\,=\,
\left[ 1+\left(\frac{\omega}{\DelZZ}\right)^{2}\right]^{-1}
\;\mbox{.}
\end{equation}
This of course is the Kondo resonance, scaling universally in terms of
the single Kondo scale $\omega_{\rm K}=\DelZZ$ but with no explicit
dependence upon the bare parameters $D$, $\DelZ$ or $U$. Its
Lorentzian shape, arising also (see eg. Ref.~\onlinecite{Hewson}) in
microscopic Fermi liquid theory and slave boson approaches, is but a
caricature of the true Kondo resonance --- which is known for example
to contain long (Doniach-\u{S}unji\'{c}) tails
\cite{Doniach70,Frota86} that are not captured by the simple
Lorentzian form. Moreover the $U$-dependence of the quasi-particle
weight $Z$, given explicitly by eq.~(\ref{Z2PT}) within 2PT, fails to
capture the exact asymptotic behaviour $Z\propto\exp(-\pi U/(8\DelZ))$
appropriate to the strong coupling Kondo regime $U/\DelZ\gg 1$,
producing instead an algebraic decay with $U$ for all
$U\gg\omo\sim\sDelZD$ as in (\ref{Z2PT_lrgU}).

The essential conclusions above are nonetheless largely independent of
2PT. Eq.~(\ref{Glowfreq}) amounts to a low-frequency expansion of the
self-energy (embodied in the first term) that is quite general,
independent of the $U$-dependence of $Z$, and valid for
$\frac{D}{\DelZZ}\abs{\omtilde}\ll 1$. This by itself is sufficient to
infer the two distinct spectral regimes, $\DelZZ\gg D$ where the
spectrum contains low-energy poles at $\omega_{-}$ (given in terms of
$\omo$ and $Z$ by eqs.~(\ref{polem_medU_Z})), and with a continuum for
$\abs{\omega}<D$ that is essentially unrenormalized by interactions;
and $\DelZZ\ll D$ where solely the continuum survives and
eq.~(\ref{Di_lrgU}) provides a reasonable description of the Kondo
resonance for $\omega/\DelZZ\ll 1$. Further, since 2PT is exact for
the two-site limit \cite{Lange98}, we anticipate that it should
provide a sound description of the observed Lanczos spectra
\cite{Hofstetter99} at least for $\DelZZ/D\gtrsim 1$, as considered
in the following section.

Before proceeding we comment briefly on the suggestion by Hofstetter
and Kehrein \cite{Hofstetter99} that the low-energy spectrum
(including the poles) in the narrow-band limit $D\to 0$, should be
described by a scaling function $\Disig(\omega)\equiv
f(\omtilde=\omega/D)$ where $f$ is independent of D.
Eq.~(\ref{Glowfreq}) supports this, since from the 2PT result
(\ref{Z2PT_lrgU}) $\DelZZ/D\sim(\DelZ/U)^{2}$ is independent of D
provided $U\gg\omo\sim\sDelZD\to 0$ (which encompasses both the $\DelZ
Z/D\gg 1$ and $\ll 1$ regimes). This however is distinct from the
one-parameter universal scaling that is characteristic of the strong
coupling Kondo regime \cite{Hewson}. The latter arises only for
$\DelZZ\ll D$, where the low-energy spectrum $\Disig(\omega)\equiv
f(\omega/\DelZZ)$ but is otherwise independent of {\em any} of the
bare parameters $D$, $\DelZ$ or $U$.

\subsection{Spectra: numerical results}
\label{sec:numspectra}

\begin{figure}[htbp]
  \begin{center} 
  \leavevmode 
  \epsfig{file=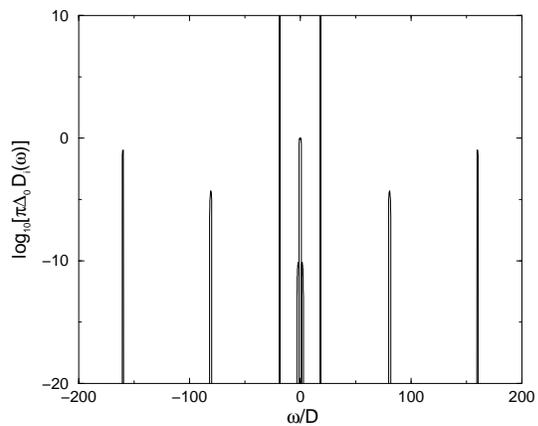,width=7cm}
  
  \caption{Low-frequency 2PT impurity spectrum, for bandwidth
  $D=10^{-4}\DelZ$ and $U=0.2\DelZ$, where $\DelZZ/D\gg 1$. Note the
  logarithmic scale for $\Disig(\omega)$, employed for clarity. The
  Hubbard satellites at $\abs{\omega}\simeq U/2=10^{3}D$ are not
  shown. Full discussion in text.}

  \label{fig:dosi} 
  \end{center}
\end{figure}

We turn now to the numerically determined 2PT
spectra. Fig.~\ref{fig:dosi} shows the low-frequency behaviour of
$\Disig(\omega)$ vs. $\omtilde=\omega/D$, for a bandwidth of
$D=10^{-4}\DelZ$ and interaction strength $U=0.2\DelZ$. For these
parameters $\DelZZ/D\simeq 540$, safely in the weak coupling regime
$\DelZZ/D\gg 1$ --- albeit 'strong' coupling in the sense that
$U/\omo\simeq 25$, as manifest in a quasi-particle weight $Z\simeq
0.054\ll 1$ that is far removed from the non-interacting limit where
$Z=1$. The spectrum is shown for $\abs{\omega}/D<200$, and we remind
the reader that almost all ($\sim 95\%$) of the spectral weight
resides in the Hubbard satellites at frequencies as large as
$U/2=10^{3} D$, which are omitted from the figure.

On the lowest energy scale $\abs{\omega}<D$ we see the expected
continuum of net weight $\order{D/\DelZ}$ which, in accordance with
the discussion above, is indeed essentially unrenormalized from the
non-interacting limit. At the band edges $\omega=\pm D$, the spectrum
vanishes of necessity since the hybridization $\ReDelta(\omega)$ is
singular there. Due to the lowest frequency self-energy continua
(\S~\ref{sec:Sigma2PT} above), the central continuum acquires side
bands ranging from $D\le\abs{\omega}\le 3 D$, whose weight remains
tiny over the whole range of interactions.

On the next higher energy scale, proportional to the antiferromagnetic
coupling $J$, we find the 'orbital remnants' at
(eq.~(\ref{omegam_medU})) $\omega_{-}=3 J/2=12 \DelZD/(\pi U)\simeq 19
D$. These poles are by far the most prominent features of the
low-energy spectrum, carrying almost all the spectral weight in this
sector, although their {\em absolute} weight
(eq.~(\ref{weight_medU_Z})) $q_{-}=Z/2\simeq 0.027$ is small since the
dominant spectral intensity is carried by the high-energy Hubbard
satellites. A third low-energy scale is the molecular bonding energy
$\omo\sim\sDelZD$ emerging from the non-interacting limit. It enters
the single-particle spectrum via the self-energy continua
(\S~\ref{sec:Sigma2PT}) and yields two sets of weak bands at
$\abs{\omega}\simeq\omo\simeq 80 D$ and $\abs{\omega}\simeq
2\omo\simeq 160 D$, albeit with net spectral weights that render them
insignificant in practice.

A comparison of Fig.~\ref{fig:dosi} with the spectrum of an
$11+1$-site Anderson star, calculated numerically for the same
parameters by Hofstetter and Kehrein \cite{Hofstetter99} with the
Lanczos method, shows that 2PT is able to reproduce essentially all
relevant energy scales. Good agreement is found in the lowest energy
scale $D$, given by the Fermi liquid continuum, and on the high-energy
scale $U/2$ for the Hubbard satellites, although 2PT remains somewhat
incomplete on the antiferromagnetic scale $J$. Here, instead of the
single pole predicted by 2PT at $\abs{\omega}=\omega_{-}\simeq 19 D$,
the Lanczos calculations show two sets of $\delta$-peaks, the first
occurring for $\abs{\omega}\simeq 18-21 D$ and the second for
$\abs{\omega}\simeq 25-26 D$. Although it is not clear how these sets
will evolve with a larger number of sites in the Lanczos calculations,
a structure with at least two features will most likely be preserved
in this region. This points to the natural limitations of 2PT, and the
need for more sophisticated many-body techniques.

For obvious numerical reasons, the Lanczos method cannot detect any
features as weak as the 2PT continua at $\omo\simeq 80 D$ and
$2\omo$. Nevertheless we believe these continua to be robust. They
cannot be cancelled by higher order self-energy diagrams since, for
any frequency, $\ImSiPT{\sigma}(\omega)\ge 0$ for {\em all}
self-energy contributions.

\begin{figure}[htbp]
  \begin{center} 
  \leavevmode 
  \epsfig{file=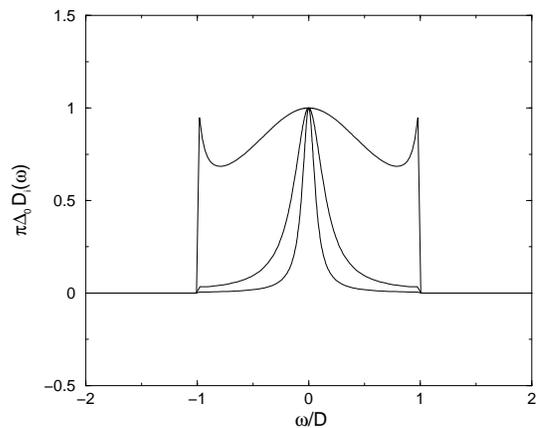,width=7cm}

  \caption{Low-frequency 2PT impurity spectrum,
  $\pi\DelZ\Disig(\omega)$ vs. $\omega/D$, for bandwidth
  $D=10^{-4}\DelZ$ and $U/\DelZ=2\pi$ (top), $4\pi$ (middle) and
  $6\pi$ (bottom), corresponding respectively to $\DelZZ/D\simeq
  0.58$, $0.15$ and $0.06$. The emergence of the Kondo resonance for
  $\DelZZ/D\ll 1$ is seen clearly.  Full discussion in text.}

  \label{fig:kondo}
  \end{center}
\end{figure}

Having discussed the weak coupling regime $\DelZZ/D\gg 1$, we now turn
to strong coupling, $\DelZZ/D\ll 1$, and the approach to
it. Fig.~\ref{fig:kondo} shows the low-energy $\Disig(\omega)$
vs. $\omtilde=\omega/D$, again for a bandwidth of $D=10^{-4}\DelZ$;
and for interaction strengths $U/\pi\DelZ=2$, $4$ and $6$,
corresponding respectively to $\DelZZ/D\simeq 0.58$, $0.15$ and
$0.06$. In agreement with the analysis of \S~\ref{sec:spectra} above,
and excluding the Hubbard satellites at $\pm U/2$, the spectra in all
cases are in practice confined exclusively to the low-frequency
continuum $\abs{\omega}<D$, whose persistent Fermi liquid character is
manifest in the pinning of the spectra at the Fermi level ($\omega=0$)
to its non-interacting value of $1/\pi\DelZ$. For the two larger
interaction strengths in particular, the Lorentzian character of the
2PT Kondo resonance arising for $\DelZZ/D\ll 1$ is seen clearly.

The results above concur with the Lanczos calculations of Hofstetter
and Kehrein \cite{Hofstetter99}, performed at a moderate interaction
strength $U/\DelZ=4$. No low-energy poles arose, and the spectrum was
found to be governed by a Fermi liquid continuum with a broad
resonance centered on the Fermi level and two sharper features at the
band edges. As shown by the top curve of Fig.~\ref{fig:kondo}, 2PT
reproduces this behaviour remarkably well, albeit that it sets in at a
somewhat larger interaction strength of $U/\DelZ\simeq 6$.

For numerical reasons, the strong coupling Kondo regime $\DelZZ/D\ll
1$ naturally cannot be captured adequately by the Lanczos calculations
\cite{Hofstetter99}, and the deficiencies of 2PT in this regime have
also been highlighted in \S~\ref{sec:spectra}. It appears however that
2PT provides a rather reasonable description of the Lanczos results
for the narrow-band AIM, and is valid up to interaction strengths
$U/\pi\DelZ$ on the order of $2$ or so, a regime of validity that
corresponds closely to that found also with 2PT for the normal
wide-band ($D=\infty$) model (see eg. Ref.~\onlinecite{Hewson}).

\section{Self-energy, and skeleton expansion.}
\label{sec:Sigma}

Within 2PT the structure of the self-energy for the narrow-band model
($\DelZ\gg D$) is rather simple, being dominated by poles at
$\omega=\pm 3\omo$, each of weight $\pi U^{2}/8$ (see
eq.~(\ref{ReSigmaPT2})). In view of the natural limitations of 2PT
however, one might ask what may be inferred more generally about the
behaviour of $\Si{\sigma}(\omega)$, and in particular pole
contributions thereto? It is aspects of this issue that we now
consider.

The self-energy is effectively defined via the Dyson equation
(\ref{Dyson}) whence, for frequencies $\omega$ where the
single-particle spectrum is purely real, $\Si{\sigma}(\omega)$ is
given by
\begin{equation}
\label{Sigma}
\Si{\sigma}(\omega)\,=\,\omega-\ReDelta(\omega)-\frac{1}{X(\omega)}
\;\mbox{.}
\end{equation}
Here $X(\omega)\equiv\real\Gisig(\omega)$ is thus defined for
convenience, and is given by the Hilbert transform
\begin{equation}
\label{ReGiHilbert}
X(\omega)
\,=\,
2\omega\int\limits_{0}^{\infty}{\rm d}\omega'
\frac{\Disig(\omega')}{\omega^{2}-\omega'^{2}}
\end{equation}
(where particle-hole symmetry has been employed, and a principal value
is implicit).
As is well known \cite{Hofstetter99} (see also
Ref.~\onlinecite{Kehrein98}), it follows from eq.~(\ref{Sigma}) that
the zeros of $X(\omega)$ correspond to the poles in
$\Si{\sigma}(\omega)$. For weak to moderate interaction strengths
($U/\DelZ\lesssim 4$), Hofstetter and Kehrein
\cite{Hofstetter99} find that the Lanczos determined $X(\omega)=0$ for
$\omega\sim\order{\omo}$ and hence that $\Si{\sigma}(\omega)$ contains
poles at frequencies on the order of $\pm\sDelZD$. This concurs
with 2PT, which we have argued (\S~\ref{sec:2PT}) to be valid in such a
weak coupling domain, where $\DelZZ/D\gtrsim 1$. But what of the
strong coupling regime $U\gg\DelZ$ ($\gg D$), where $\DelZZ/D\ll 1$
and the Kondo effect is well developed?
The limitations of the Lanczos technique in this domain have been
alluded to above, but the general structure of the Lanczos-determined
spectrum $\Disig(\omega)$ is in fact sufficient to determine the poles
of $\Si{\sigma}(\omega)$, as now shown.

In strong coupling the single-particle spectrum is found
\cite{Hofstetter99} to separate into high- and low-energy parts,
$\Disig(\omega)=\Disig^{\rm H}(\omega)+\Disig^{\rm L}(\omega)$; we
focus below on $\omega>0$ (by particle-hole symmetry). The high-energy
contribution is $\Disig^{\rm H}(\omega)=
q_{+}\delta(\omega-\omega_{+})$.  It corresponds to the Hubbard
satellite, with position $\omega_{+}\sim U/2$ and pole-weight
$q_{+}\sim 1/2$. By contrast the low-energy continuum $\Disig^{\rm
L}(\omega)$ is non-zero only on the band scale $\abs{\omega}\le D$, is
pinned at the Fermi level where $\pi\DelZ\Disig(\omega=0)=1$
ubiquitously, and contains a well developed Kondo resonance on the
scale $\omega\sim\DelZZ\ll D$. From eq.~(\ref{ReGiHilbert}) it
follows that $X(\omega)=X_{\rm H}(\omega)+X_{\rm L}(\omega)$, given
asymptotically by
\begin{mathletters}
\label{ReGiAsym}
\begin{eqnarray}
\label{ReGiHAsym}
X_{\rm H}(\omega) 
&\stackrel{\omega\ll U/2}{\sim}&
-\frac{2 q_{+}}{\omega_{+}^{2}} \omega
\, \sim \,
-\frac{4}{U^{2}} \omega
\\
\label{ReGiLAsym}
X_{\rm L}(\omega) 
&\stackrel{\omega\gg D}{\sim}&
\frac{\delta}{\omega}
\end{eqnarray}
\end{mathletters}
where $\delta=\int_{-D}^{+D}{\rm d}\omega \Disig^{\rm L}(\omega)$ is
the integrated weight of the low-energy continuum. From
eq.~(\ref{ReGiAsym}) the zeros of $X(\omega)$, and hence the poles of
$\Si{\sigma}(\omega)$, thus occur at $\omega=\pm\omega_{P}$ given by
\begin{equation}
\label{omegaSiPole1}
\omega_{P}\,=\,\omega_{+}\sqrt{\frac{\delta}{2 q_{+}}}
\,\sim\,
\frac{U}{2}\sqrt{\delta}
\end{equation}
(which is self-consistent provided $D\ll\omega_{P}\ll U/2$ as indeed
found, see below); the corresponding pole-weight in
$\Si{\sigma}(\omega)$, given from eq.~(\ref{Sigma}) by
$Q=\pi/\abs{\partial X(\omega)/\partial\omega}_{\omega=\omega_{P}}$,
is
\begin{mathletters}
\label{poleSiPole}
\begin{equation}
\label{weightSiPole}
Q
\,=\,
\frac{\pi\omega_{+}^{2}}{4 q_{+}}
\,\sim\,
\frac{\pi U^{2}}{8}
\;\mbox{.}
\end{equation}
Since $\Disig(\omega)$ is normalized to unity however, the integrated
low-energy spectral weight $\delta=1-2 q_{+}$. In strong coupling
moreover, where $2\abs{\varepsilon_{i}}=U\gg\DelZ$ ($\gg D$), the
leading asymptotic behaviour of the high-energy spectral pole-weight
$q_{+}$ is in turn readily deduced by taking the limit of constant
$V_{i\vk}$ and $D\to 0$, {\sl i.e.} by considering fixed
$\omo\sim\sDelZD$ and $D\to 0$, which is just the two-site limit for
which \cite{Lange98} $q_{+}\sim 1/2-18\omo^{2}/U^{2}$ (amounting in
effect to perturbation theory in $V_{i\vk}/U$). Hence $\delta\sim
36\omo^{2}/U^{2}$, and from eq.~(\ref{omegaSiPole1}),
\begin{equation}
\label{omegaSiPole}
\omega_{P}\,=\,3\omo
\;\mbox{.}
\end{equation}
\end{mathletters}

Eq.~(\ref{omegaSiPole}) (with $Q$ from (\ref{weightSiPole})) is the
result in strong coupling $U\gg\DelZ$ for the poles in
$\Si{\sigma}(\omega)$. The arguments for it do not depend upon
perturbation theory in $U$ about the non-interacting limit, but the
result itself is precisely that arising from simple 2PT,
eq.~(\ref{ReSigmaPT2}). This is salutory, for although 2PT
extrapolated to strong coupling produces an inadequate caricature of
the Kondo resonance on scales $\omega\sim\order{\DelZZ}\ll D$ (see
e.g. \S~\ref{sec:spectra}), the above arguments suggest that on scales
of order $\omo\sim\sDelZD$ it becomes asymptotically exact in strong
coupling. In addition, noting that $\omo$ is $U$-independent and that
eq.~(\ref{omegaSiPole}) holds both perturbatively for
$U\lesssim\order{\DelZ}$ and in strong coupling $U\gg\DelZ$, it is
clear that any dependence of the poles in $\Si{\sigma}(\omega)$ upon
$U$ is at best weak across the entire range of interaction strengths.

\subsection{Skeleton expansion}
\label{sec:skeleton}

There is a second feature of the above analysis to which we would draw
attention, namely that the poles in $\Si{\sigma}(\omega)$ connect
continuously to that characteristic of the atomic limit, $\DelZ=0$
({\sl i.e.} $V_{i\vk}=0$). As $\DelZ\to 0$ the two poles in
$\Si{\sigma}(\omega)$, each of weight $Q=\pi U^{2}/8$ and occurring at
$\omega=\pm 3\omo$ with $\omo\sim\sDelZD$, coalesce to a single pole
at $\omega=0$ with weight $\pi U^{2}/4$, thus producing
\begin{equation}
\label{SigmaAtomic}
\Si{\sigma}(\omega)\,=\,\frac{U^2}{4}\frac{1}{\omega+\iz\sign{\omega}}
\qquad:\;\DelZ=0
\;\mbox{.}
\end{equation}
This result, which corresponds to 2PT, is of course well known
\cite{Hewson} to be exact in the atomic limit for all $U>0$, a
feature that is particular to the particle-hole symmetric case considered.

Hofstetter and Kehrein \cite{Hofstetter99} have considered the
behaviour of $\Si{\sigma}(\omega)$ for the narrow-band AIM from the
perspective of the skeleton expansion, where the {\em exact}
propagator $\Gisig(\omega)$ is inserted into every skeleton diagram
contributing to $\Si{\sigma}(\omega)$. They point out that the pinning
of the single-particle spectrum at the Fermi level (for all $U\ge 0$
and any $\DelZ\ge 0$) attests to the convergence of the skeleton
expansion on the lowest energy scales characteristic of the Fermi
liquid continuum. Conversely, following Kehrein \cite{Kehrein98}, they
show that the $\abs{\omega}\sim\order{\omo}$ poles in
$\Si{\sigma}(\omega)$ cannot be explained in any order of the skeleton
expansion, which fails to converge for such frequencies. Given the
above connection between the poles in $\Si{\sigma}(\omega)$ and that
endemic to the atomic limit, the latter behaviour is entirely natural,
since it is known that the skeleton expansion fails to converge for
the atomic limit (and more generally, but in essence equivalently, for
an insulator \cite{Kehrein98}).

In view of the above it may be instructive to consider the failure of
the skeleton expansion, taking the atomic limit as a paradigm. The
essence of the skeleton expansion to any finite order is that it
involves a {\em partial} infinite-order summation of diagrams obtained
from PT in $U$ (and themselves expressed in terms of the
non-interacting propagator $\GiFsig(\omega)$). As such however, it may
fail to include higher-order diagrams that act in large part to cancel
those that are included, and this is where the dangers arise.

Consider for example the following diagram, which is 4th order in PT
but is included in the class of 2nd order skeletons:
\begin{equation}
  \label{fig:SigmaSkel2}
  \leavevmode
  \unitlength=2ex 
  \setlength{\Units}{\the\unitlength} 
  \begin{picture}(6,8)
  \put(0 ,-1.0){\epsfig{file=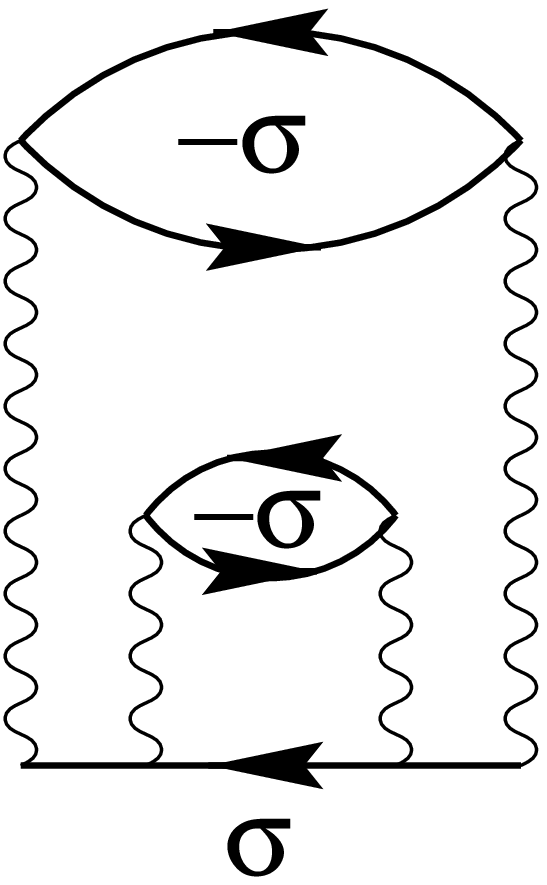,width=5\Units}}
  \end{picture}
\end{equation}
It represents a process creating two particle-hole excitations of spin
$-\sigma$ on the impurity site. If the first such pair has not hopped
off the impurity when the second pair is created --- as is inexorable
in the atomic limit --- then such processes are formally forbidden by
the Pauli principle. They are nonetheless properly cancelled by
exchange diagrams arising from the same order in PT, for example:
\begin{equation}
  \label{fig:SigmaSkel2x}
  \leavevmode
  \unitlength=2ex 
  \setlength{\Units}{\the\unitlength} 
  \begin{picture}(6,8)
  \put(0 ,-1.0){\epsfig{file=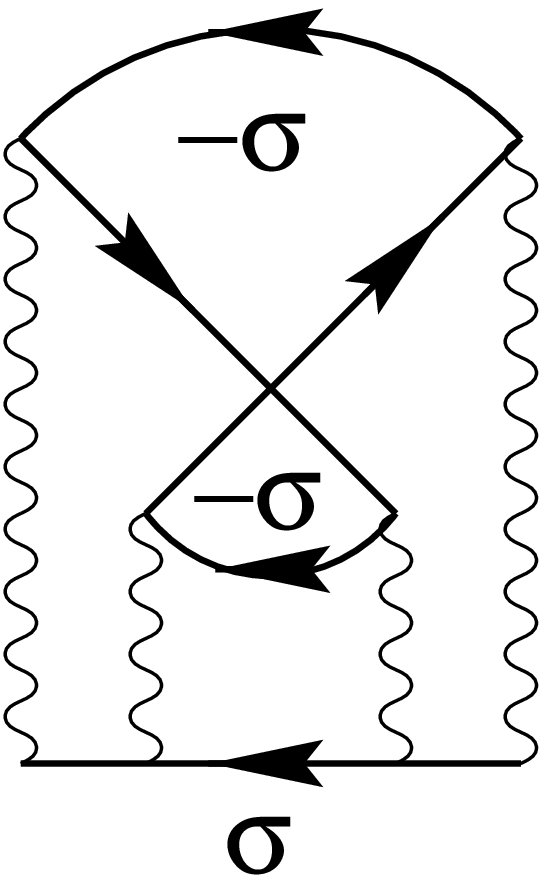,width=5\Units}}
  \end{picture}
\end{equation} 
However the exchange diagram (\ref{fig:SigmaSkel2x}), while arising to
the same order in PT as (\ref{fig:SigmaSkel2}), belongs to the class
of 4th order skeletons. In this sense it is of higher order, and would
not therefore be included if the skeleton expansion was truncated at
2nd order.

Generalizing this reasoning illustrates that the skeleton expansion,
if truncated after a certain class, will include specific series of
diagrams without necessarily accounting for the corresponding exchange
terms, since the latter belong in general to a higher skeleton
class. Hence, even for systems where PT in $U$ is known to converge
order-by-order, the order-by-order convergence in {\em skeletons} may
not be taken for granted.

The atomic limit, whose relevance to the narrow-band AIM has been
pointed out above, provides a direct example of the latter point. Here
$\Si{\sigma}(\omega)$ is given exactly by 2PT
(eq.~(\ref{SigmaAtomic})), {\sl i.e.} by
$\Si{\sigma}(\omega)=\frac{U^2}{4}\GiFsig(\omega)$ (where, trivially,
the unperturbed $\GiFsig(\omega)=1/(\omega+\iz\sign{\omega})$. The
problem is convergent order-by-order in straight PT, all diagrams
contributing to $\Si{\sigma}(\omega)$ in any given order $n>2$ of PT
summing precisely to zero. Consider by contrast the second-order
skeleton expansion, denoted $\Sigma_{\rm skel}^{(2)}$ and obtained
from eq.~(\ref{fig:Sigma2PT}) by replacing $\GiFsig$ with the exact
$\Gisig$, to give \cite{Lange98}
$
\Sigma_{\rm skel}^{(2)}(\omega)
=\frac{U^{2}}{4}\omega/[\omega^{2}-(3 U/2)^{2}]
\stackrel{\omega\to 0}{\sim} -\omega/9
$.
This fails entirely to capture the $\sim 1/\omega$ behaviour of the
exact $\Si{\sigma}(\omega)$ \cite{Lange98} (save trivially for
$\abs{\omega}\gg 3 U/2$), and that it does so for the essential
reasons outlined above is directly evident by recasting $\Sigma_{\rm
skel}^{(2)}$ as a functional of $\GiFsig$, viz:
\begin{equation}
\label{SigmaSkel2}
\Sigma_{\rm skel}^{(2)}(\omega)
\;=\;\frac{\frac{U^{2}}{4}\GiFsig(\omega)}
{1\,-\,\left[\frac{3 U}{2}\GiFsig(\omega)\right]^{2}} 
\end{equation}
 
Finally, we note that the behaviour described in this section is not
specific to the narrow-band AIM, but arises also in the
infinite-dimensional Hubbard model at half filling (in which context
study of the narrow-band AIM was first motivated \cite{Hofstetter99}).
Within IPT \cite{Georges96,Georges92,IPT}, as $U$ approaches the
critical $U_{c}$ for the Mott transition from the metallic phase
$U<U_{c}$, the self-energy acquires poles (or strictly, sharp
resonances) at finite $\omega$ in a preformed gap; and as $U\to
U_{c}-$ the pole positions approach the Fermi level $\omega=0$ (less
rapidly than the central Fermi liquid continuum vanishes), and with
weights that remain finite. Hence as $U\to U_{c}-$ the low-frequency
behaviour characteristic of an insulator --- $\Si{\sigma}(\omega)\sim
A/(\omega+\iz\sign{\omega})$ which amounts in essence to that for the
atomic limit --- is smoothly recovered via IPT. The poles in
$\Si{\sigma}(\omega)$ are thus an integral facet of the Mott
transition and, from the discussion above, we believe the success of
IPT in capturing it is intimately connected to its ability
\cite{Georges96,Georges92,IPT} to recover correctly the atomic
limit. By contrast, since the poles in $\Si{\sigma}(\omega)$ cannot be
captured via the skeleton expansion \cite{Kehrein98}, approaches based
at heart upon the latter --- such as self-consistent perturbation
theory \cite{MuellerHartmann} --- simply fail to uncover the
transition \cite{Lange98}.

\section{Conclusions} 

We have studied in this article the symmetric AIM with a narrow host
band, $D\ll\DelZ$. Simple though it is, second order perturbation
theory in $U$ is found to give a rich and relatively complete account
of the underlying single-particle dynamics. In particular, and in
agreement with recent Lanczos calculations \cite{Hofstetter99}, it
leads naturally to two distinct regimes of spectral behaviour
according to whether $\DelZZ/D\gg 1$ or $\ll 1$, and whose essential
characteristics we have argued to be largely independent of the
details of 2PT itself.

We have also shown that 2PT is remarkably robust on energy scales of
order $\abs{\omega}\sim\sDelZD$, that reflect the underlying molecular
orbitals characteristic of the non-interacting limit. Here we have
argued that the 2PT result for the self-energy is correct both
perturbatively in weak coupling (by construction), {\em and} in the
strong coupling limit $U\gg\DelZ$. Such poles in $\Si{\sigma}(\omega)$
--- which cannot be captured to any order in a skeleton expansion
\cite{Hofstetter99,Kehrein98} --- thus appear to be an essentially
$U$-independent characteristic of the narrow-band AIM, and that they
are captured by 2PT is in turn closely related to its ability to
recover correctly the atomic limit of the model.

The latter comments also illustrate the difficulties that confront any
theory which seeks to describe single-particle dynamics of the
narrow-band AIM in strong coupling, and on all energy scales. At low
energies $\omega\sim\omega_{\rm K}\ll D$ characteristic of the Kondo
resonance, low-order PT in $U$ will naturally not suffice and an
intrinsically non-perturbative approach will be needed. And yet for
energy scales on the order of $\sDelZD\ll\DelZ$ the essential result
of 2PT must be recovered --- which as we have discussed is a delicate
matter, since 'obvious' approaches based on partial infinite-order
summation of PT in $U$ are liable to qualitative failure here.  While
accessible to numerical techniques such as the numerical
renormalization group \cite{Hewson}, we do not know of any
conventional theoretical approach that can encompass these twin
dictates. In a subsequent paper we will however show that a recently
developed local moment approach \cite{Logan98,Bulla00} can handle the
problem.

One of the authors (S.~Sch\"afer) acknowledges a postdoctoral
fellowship of the Deutscher Akademischer Austauschdienst (DAAD) under
grant no.~D/98/27069.  
DEL is grateful to the EPSRC for financial support.

% %%%%%%%%%%%%%%%%%%%%%%%%%%%%%%%%%%%%%%%%%%%%%%%%%%%%%%%%%%%%
% REFERENCES
% %%%%%%%%%%%%%%%%%%%%%%%%%%%%%%%%%%%%%%%%%%%%%%%%%%%%%%%%%%%%

% %%%%%%%%%%%%%%%%%%%%%%%%%%%%%%%%%%%%%%%%%%%%%%%%%%%%%%%%%%%%
% FIGURES
% %%%%%%%%%%%%%%%%%%%%%%%%%%%%%%%%%%%%%%%%%%%%%%%%%%%%%%%%%%%%

\newpage

\end{multicols} 

\end{document}